\documentclass[prb,twocolumn,showpacs,preprintnumbers,amsmath,amssymb,superscriptaddress,footinbib]{revtex4-1}
\usepackage[export]{adjustbox}
\usepackage{amsfonts,amsmath,times,mathrsfs,graphicx,dcolumn,bm,wasysym,xcolor,epsfig,amssymb,lipsum,color,epstopdf,bbold,booktabs,bbm,mathtools}
\usepackage[makeroom]{cancel}
\usepackage[english]{babel}
\usepackage[T1]{fontenc}
\usepackage[colorlinks,bookmarks=true,citecolor=red,linkcolor=blue,urlcolor=blue]{hyperref}
\usepackage{hyperref}
\usepackage[tight, FIGTOPCAP, hang, raggedright, nooneline]{subfigure}
\usepackage{xcolor}
\usepackage{physics}
\definecolor{Green}{RGB}{80,182,0} 
\usepackage{float}
\usepackage{amsmath}

\begin{document}
\setlength{\parskip}{0pt}

\title{Homogeneous Floquet time crystal from weak ergodicity breaking}

\author{H. Yarloo}
\email{yarloohadi@gmail.com}
\affiliation{Department of Physics, Sharif University of Technology, P.O.Box 11155-9161, Tehran, Iran}

\author{A. Emami Kopaei}
\affiliation{Department of Physics, Sharif University of Technology, P.O.Box 11155-9161, Tehran, Iran}

\author{A. Langari}
\email{langari@sharif.edu}
\affiliation{Department of Physics, Sharif University of Technology, P.O.Box 11155-9161, Tehran, Iran}

\begin{abstract}
Recent works on the observation of discrete time-crystalline signatures throw up major puzzles on the necessity of localization for stabilizing such out-of-equilibrium phases. Motivated by these studies, we delve into a clean interacting Floquet system, whose quasi-spectrum conforms to the ergodic Wigner-Dyson distribution, yet with an unexpectedly robust, long-lived time-crystalline dynamics in the absence of disorder or fine-tuning.
We relate such behavior to a measure zero set of nonthermal Floquet eigenstates
with long-range spatial correlations, which coexist with otherwise thermal states at  near-infinite temperature and develop a high overlap with a family of translationally invariant, symmetry-broken initial conditions. 
This resembles the notion of ``dynamical scars'' 
that remain robustly localized throughout a thermalizing Floquet spectrum with fractured structure. We dub such a long-lived discrete time crystal formed in partially nonergodic systems, ``scarred discrete time crystal'' which is distinct by nature from those stabilized by either many-body localization or prethermalization mechanism.
\end{abstract}
\maketitle
\section{Introduction}\label{sec:intro}
Periodically driven (Floquet) quantum systems are of immense recent interest as they can sustain a variety of novel solid state phenomena ranging from Floquet engineering~\cite{Eckardt:2017,Bukov:2015_rev,Oka:2019}
to extending the theory of localization
or Mott insulators
to the time domain~\cite{Sacha:2015_ext,Sacha:2016_AL,Mierzejewski:2017,Delande:2017}. They also provide natural platforms for realizing intriguing topological phases,
hosting anomalous chiral edge states~\cite{Rudner:2013,Titum:2016,Mukherjee:2017}
or Majorana edge modes~\cite{Jiang:2011,Thakurathi:2013},
as well as emergent non-equilibrium phases of matter with no static equilibrium counterpart. One of the most significant  phases is Floquet discrete time-crystal (DTC)~\cite{Sacha:2015,Else:2016}, the so-called ``$ \pi $ spin glass''~\cite{Khemani:2016_PRL,Keyserlingk:2016,Keyserlingk:2016SSB}, in which a driven system fails to be invariant under the discrete time-translation symmetry of its underlying Hamiltonian.

More broadly, the concept of time crystal has to do with the spontaneous emergence of time-translation symmetry breaking (TTSB) within a time-invariant system. 
In  2012, Wilczek conceptualized the possibility of continuous TTSB 
for the ground state of a certain quantum and classical system~\cite{Wilczek:2012,Shapere:2012}. 
However, his original proposition has triggered an intense debate~\cite{Bruno:2013_comment,Wilczek:reply,Li:2012}, including subsequent no-go theorems~\cite{Bruno:2013,Philippe:2013,Watanabe:2015}, concerning the existence of time crystals at thermal equilibrium and in the ground states of local time-independent Hamiltonians.
On this basis, the search for time crystals shifted toward certain nonequilibrium conditions~\cite{Syrwid:2017}, 
in particular, Floquet systems~\cite{Sacha:2015,Else:2016,Khemani:2016_PRL}. 
The defining diagnostic of a stable DTC phase then reads as non-trivial subharmonic response of certain physical observables, at some multiple of the drive frequency, which is robust to generic perturbations and persists infinitely on approaching the thermodynamic limit.

Nevertheless, such a discernible phase structure is generically nonviable so long as the strong form of ``eigenstate thermalization hypothesis'' (ETH)~\cite{Jensen:1985,Deutsch:1991,Srednicki:1994,Rigol:2008} holds in that \textit{all} Floquet eigenstates look like maximally entangled featureless states~\cite{Alessio:2014,Lazarides:2014}. 
Thus the key strategy for stabilizing temporal order  is to explore possible ways to completely suppress (or at least slow down) the process of Floquet heating toward infinite temperature. 
This can typically be achieved by either considering (fine-tuned) Bethe-ansatz integrable systems~\cite{Sutherland:2004}, 
or extending the physics of many-body localization (MBL)~\cite{Basko:2006,Oganesyan:2007}, driven by spatial disorder, to the Floquet realm~\cite{Ponte:2015}.
The latter provides the only known generic mechanism for strong breakdown of ergodicity due to the emergence of a complete set of quasi-local integrals of motion
in the so-called ``l-bits'' formalism~\cite{Huse:2014,Serbyn:2013_2}. 
The fully localized spectrum of Floquet-MBL systems can establish spatio-temporal order 
even at infinite temperature, 
giving rise to the concrete
example of \textit{absolutely stable} (space-)time crystals~\cite{Else:2016,Yao:2017,Khemani:2016_PRL,Keyserlingk:2016,Zhang:2017,Smits:2018}. 
However, MBL is not the only game in town. 
So far, a range of mechanisms have been exploited,
both theoretically and experimentally, to realize robust DTC phase (or at least transient DTC signatures) in a broad class of generic clean systems~\cite{Bukov:2016,Abanin:2017,Else:2017-b,Luitz:2019,Rovny:2018,Peng:2019,Machado:2020,Zeng:2017_PDTC,Huang:2018,Mizuta:2018,Iadecola:2018,Russomanno:2019,Chinzei:2020,Sacha:2015,Russomanno:2017,Nurwantoro:2019,Pizzi:2019,Matus:2019,Pal:2018,Barfknecht:2019,Gong:2018,Luitz:2018,Schafer:2019}. 
These mechanisms go from 
prethermalization~\cite{Bukov:2016,Abanin:2017,Else:2017-b,Luitz:2019,Rovny:2018,Peng:2019,Machado:2020,Zeng:2017_PDTC} to emergent Floquet integrability in systems with strong interactions~\cite{Huang:2018}, as well as those relied on a protecting ``ancillary'' symmetry~\cite{Else:2019}, e.g., spatial translation~\cite{Mizuta:2018}, time-reflection~\cite{Iadecola:2018}, or discrete (Abelian) gauge symmetry~\cite{Russomanno:2019}.

A rather crisp realization of DTC is also provided by periodically driven mean-field models~\cite{Sacha:2015,Russomanno:2017,Nurwantoro:2019,Pizzi:2019,Matus:2019,Pal:2018,Barfknecht:2019,Gong:2018}, which can exhibit discrete TTSB 
even in the presence of \textit{quantum chaos}. 
The realization of this kind of DTC, however, 
is tied to an intrinsically semiclassical \textit{few-body} phenomenon
rather than quantum many-body interactions, whose presence is essential for stabilizing MBL-driven and prethermal DTCs.
Such an exotic behavior can be attributed to the 
phenomenon of \textit{mixed} classical phase space (and its semiclassical correspondence for quantum few-body systems~\cite{Haake:2018,Percival:1973,Reichl:1987,Bohigas:1993}), in which chains of regular ``islands'' are surrounded by a chaotic ``sea''. 
The periodic jump among separated islands leaves DTC imprint on quantum dynamics when the initial state predominantly falls
inside 
one of these regular regions~\cite{Russomanno:2017,Nurwantoro:2019,Pizzi:2019}. 
The rigidity of 
mean-filed time crystals then owes to
the stability of 
the mixed phase space under 
weak integrability-breaking perturbations, 
which is ensured by the Kolmogorov-Arnold-Moser (KAM) theorem~\cite{Kolmogorov:1954}. 
However, away from semiclassical limit for many degrees of freedom, the conditions of the KAM theorem become fragile
and one expects the quick disappearance of regular islands,  which turns the system into a trivial ETH phase~\cite{Russomanno:2015,Russomanno:2017,Nurwantoro:2019,Pizzi:2019}.

By contrast, here we aim to investigate the formation of robust time-crystalline order, beyond the  semiclassical limit, in a generically chaotic \textit{many-body} system as a consequence of weak ergodicity breaking~\cite{Biroli:2010,Banuls:2011,Ikeda:2013,Kim:2014,Shiraishi:2017,Kormos:2016,Robinson:2018}. 
The weak form of ETH allows for the existence of a measure zero set of ETH-violating eigenstates at finite energy density, which are embedded 
in a sea of thermalizing states and now named ``quantum many-body scars''~\cite{Turner:2018,Ho:2019}. 
The scar states, then by definition, can evade the prescribed no-go arguments of Ref.~\onlinecite{Watanabe:2015}, which in turn  allows for TTSB-like behavior in quench dynamics of certain kinetically constrained models~\cite{Bernien:2017,Turner:2018,Ho:2019,Schecter:2019,Iadecola:2019_magnon,Mukherjee:2019}. 
However, the perfect scars and the resulting TTSB 
typically are limited to rather \textit{fine-tuned} 
settings~\cite{Shiraishi:2017,Choi:2019_SU2,Ok:2019,Bull:2019,Moudgalya:2018,Moudgalya:2018_AKLT,Schecter:2019,Iadecola:2019_magnon,Chattopadhyay:2020,Mukherjee:2019,Haldar:2019,Zhao:2020}, and not expected to be robust under generic perturbations~\cite{Khemani:2019,Lin:2019}.

It has been recently argued that a more \textit{robust} types of scars, 
and hence ergodicity breaking, can arise from ``Hilbert-space fragmentation''~\cite{Sala:2020,Khemani:2019_shatter,Pai:2019,Pai:2019_fractonlocalization,Hudomal:2020,Yang:2019,Pai:2020_con,Tomasi:2019_frac,Moudgalya:2019,Morningstar:2020}, where the Hilbert space fractures into exponentially many finite or even infinite~\cite{Moudgalya:2019} size Krylov subspaces that remain dynamically disconnected (closed) even after resolving all possible explicit symmetry sectors of the Hamiltonian. The dynamical fracturing leads to an initial-state dependent, effectively localized dynamics that stands beyond the scope of locator-expansion techniques. 
The most promising candidate in this direction is fractonic Floquet
random circuit models~\cite{Pai:2019,Pai:2019_fractonlocalization,Khemani:2019_shatter}, 
in which a subset of robust, localized steady states manifest in the thermalizing Floquet spectrum  independent of microscopic details of circuits or driving protocols.
These atypical eigenstates, characterized by the subthermal entanglement, are referred to as ``dynamical scars''~\cite{Pai:2019} in analogy to their static counterparts. 
It is therefore natural to ask whether such \textit{partially nonergodic phases} can open the door to exploring robust time-crystalline behavior in the presence of many-body quantum chaos?
\begin{figure*}[t!]
\centering
\includegraphics[width=0.965\linewidth]{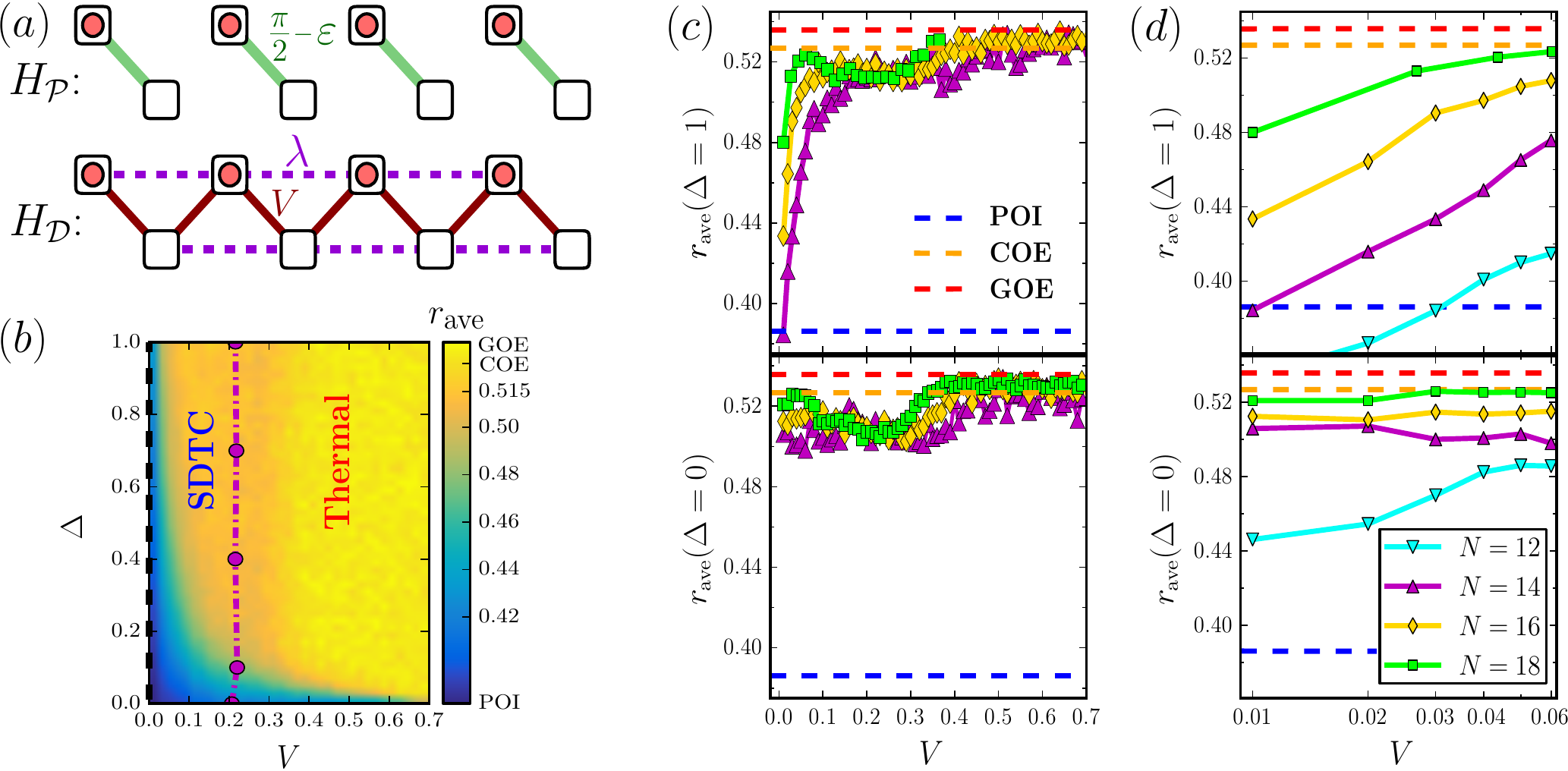}
\caption{(a) Illustrative representation of the Floquet Hamiltonian~\eqref{eq:FM1}. (b) Color plot of the level statistics ratio $r_\mathrm{ave}$ in the plane of $ {(V,\Delta)} $ at fixed ${\lambda=0.7}$ for an open chain with ${N=16}$.  Circles indicate the crossover between SDTC and a trivial thermalizing regimes. 
The crossover is obtained from the analysis of the dynamics of local imbalance autocorrelators using confused recurrent neural network in Fig.~\ref{fig:RNN_confusion}.
(c) Finite-size behavior of $r_\mathrm{ave}(V,\Delta)$, shown in (b), 
at fixed finite value of ${\Delta=1}$ (top panel), and the inversion symmetric line ${\Delta=0}$ (bottom panel). 
For 
the latter case, the results are averaged over 
different symmetry sectors. Panel (d) displays the integrable looking region near ${V=0}$  
at fixed $ \Delta $, 
where $r_\mathrm{ave}$ flows with system size to the random matrix prediction. 
}
\label{fig:MLS_OBC_mesh_line} 
\end{figure*}

To answer this question, we begin with a simple  nonintegrable Floquet model, as a concrete example of clean true DTCs stabilized by emergent Floquet integrability~\cite{Huang:2018}.
We then show that an infinitesimal deformation of this model is enough to completely destroy signatures of integrability in the Floquet spectrum and generates quantum chaos in the thermodynamic limit.
However, the system still features robust, long-lived discrete TTSB which does not hinge upon either of fine-tuning or disorder and rather stabilized by long-range correlated, dynamical scar states;  hence the name ``\textit{scarred discrete time crystal}'' (SDTC).

The appearance of such an exotic behavior is remarkable in light of the lack of any protecting ancillary symmetry (e.g., time-reflection~\cite{Iadecola:2018}) or \textit{explicit} local constraint (e.g., 
fracton-like constraints~\cite{Sala:2020,Khemani:2019_shatter,Pai:2019,Pai:2019_fractonlocalization,Moudgalya:2019}) that impedes dynamical scars from mixing with typical thermal states. 
The rigidity of the SDTC can then be understood through the stability of dynamical scars under generic perturbations.
Using \textit{confused} recurrent neural network, 
as a semisupervised machine learning method, we 
affirmatively 
confirm the robustness of the SDTC response in the limit of strong interactions, beyond which the system will eventually thermalize (see Fig.~\ref{fig:MLS_OBC_mesh_line}(b)). 
We also explain the formation of the SDTC through the emergence of long-lived, local quasi-conservation laws in the form of \textit{state-dependent},  
local integrals of motion.
Once the system is properly initialized, the SDTC dynamics cannot strikingly evolve out of the underlying initial sector and approximately preserves 
the conservation laws in question. 
Our study thus suggests the existence of a new class of time crystals which is neither localization-driven (via MBL mechanism
or gauge invariance)
nor symmetry protected,
and remarkably would be the case even if the standard prethermal mechanism or mean-field treatment
is inapplicable.

The paper is structured as follows. In Sec.~\ref{sec:model} we introduce the model and describe its relation with the emergent Floquet integrable model found in  Ref.~\onlinecite{Huang:2018}.
Section.~\ref{sec:TTSB} describes in detail 
our findings regarding the spectral statistics and structure of Hilbert space from the point of view of 
the ETH, quantum correlation and entanglement.
In Sec.~\ref{sec:PRC} we address the persistence, initial-state dependence, and rigidity of the SDTC. 
Section.~\ref{sec:WEB} describes the emergence of local quasi-conservation laws in the SDTC regime. 
We conclude the paper by briefly summarizing our main results with discussions in Sec.~\ref{sec:Discussion}.
\section{the model}\label{sec:model}
We begin by considering a clean one dimensional lattice model of interacting spinless fermions, which undergoes a periodic driving dictated by Floquet unitaries of the form ${\mathcal{U}_F= \hat{\mathcal{P}} e^{-it_\mathcal{D} H_\mathcal{D}}}$, where ${ \hat{\mathcal{P}}=e^{-it_\mathcal{P} H_\mathcal{P}}} $ (${\hbar=1}$) and, 
 \begin{align} \label{eq:FM1}
 H_\mathcal{P} =& (\frac{\pi}{2}-\varepsilon)\sum_{i \in \text{odd}}^N \hat c^\dagger_{i+1} \hat c_{i} + \mathrm{H.c.},\\
 H_\mathcal{D} =& V \sum_i (\hat c^\dagger_{i+1} \hat c_{i}+ \mathrm{H.c.})+\lambda \sum_{\ll ij \gg} \hat n_{i} \hat n_{j}+\Delta \sum_i (-1)^i \hat n_i\nonumber,
 \end{align}
are the two portions of binary stroboscopic Floquet Hamiltonian during $ t_\mathcal{P} $ and $ t_\mathcal{D} $, respectively (see Fig.~\ref{fig:MLS_OBC_mesh_line}(a)). 
Here, $\hat c_{i} $ 
and ${\hat n_{i}= \hat c^{\dagger}_{i} \hat c_{i}} $, respectively, represent the annihilation 
and local occupation operator of fermions at site $ i $, $N$ denotes the number of lattice sites at half-filling, and  $ {T=t_\mathcal{P}+t_\mathcal{D}} $ is the drive period. 
To simplify the notation, the coupling constants ($ \varepsilon,\lambda,V,\Delta $) measure in the unit ${t_\mathcal{P} = t_\mathcal{D} = 1}$, so ${\omega_0 = {2\pi}/{T}= \pi}$.

It has been recently shown that this model with ${V=0}$, denoted by ${\mathcal{U}_{F}^{int} \equiv \mathcal{U}_{F}(\varepsilon,\lambda,\Delta)}$, can feature genuine time-crystal order protected by emergent Floquet integrability~\cite{Huang:2018}. 
Accordingly, 
we first sketch the dynamics governed by stroboscopic time evolution operator, ${U(nT)=(\mathcal{U}_{F}^{int})^n}$.  
For $ {\varepsilon=0} $, the pumping term $ \hat {\mathcal{P}} $ perfectly exchange particles between even and odd sites regardless of the driving field $ H_{\mathcal{D}} $.  
Hence the local fermion imbalance between even and odd lattice sites of the $ i^{\mathrm{th}} $ unit cell, denoted by ${ \hat{\mathcal{I}}_i=\hat n^e_{i}-\hat n^o_{i}} $, changes its sign once per Floquet period. 
Consequently, measuring the temporal autocorrelation function ${\langle \hat{\mathcal{I}}_i(nT) \hat{\mathcal{I}}_i(0)\rangle} $ at stroboscopic times leads to $ 2T $-Rabi oscillations.
In the single-particle limit $ {\lambda=0} $, such a temporal order is unstable to generic imperfection, $ {\varepsilon\neq0} $. 
However, in the limit of strong interaction $ {\lambda/ \varepsilon\gg1} $, $ H_{\mathcal{D}} $ can act as a collective synchronizer and cause the period to be spontaneously doubled.
This coherent dynamics owes to the presence of an ``incomplete'' set of emergent 
Floquet local integrals of motion,
which cause the whole spectrum of ${\mathcal{U}_{F}^{int}}$ to harbor uncorrelated quasi-energy levels characterized by 
``imperfect'' 
Poisson statistics~\cite{Huang:2018}.  
However, this emergent integrability is not exact and the distribution of level spacings is close to (and not quite) Poisson as $ {N\to\infty} $.

Despite being in a \textit{finite} distance from an  \textit{exact} (emergent)
integrable manifold, it has been claimed that 
this model 
does realize 
\textit{true} DTC phase with an exponentially diverging lifetime in system size~\cite{Huang:2018}. 
Building upon this work, we rule out the formation of such  genuine 
temporal order 
as it is \textit{not} absolutely stable (at least) against symmetry-breaking perturbations, e.g., $V$ term in \eqref{eq:FM1}, and hence is not generic. 
As will be shown below, adding an infinitesimal $ V $ perturbation,  even at $ {\varepsilon=0} $, 
substantially modifies the spectral statistics of  ${\mathcal{U}_{F}^{int}}$, making the model \textit{generically chaotic} for $ {N\to\infty} $
(see  Figs.~\ref{fig:MLS_OBC_mesh_line}(b) to~\ref{fig:MLS_OBC_mesh_line}(d)). 
In this way, 
the signatures of Floquet integrability and the 
\textit{seemingly} true DTC order in ${\mathcal{U}_{F}^{int}}$ 
are somewhat fine-tuned
to a certain manifold at $ {V=0} $. 
Nevertheless, we still observe 
long-lived subharmonic oscillations in the dynamics of the 
chaotic deformed model ${\mathcal{U}_{F}}$ (see e.g., Fig.~\ref{fig:iTEBD_m}(a)).

Hereunder, we set ${\varepsilon = 0}$ and use the drive imperfection $V$, which leads to beating in ${\langle \hat{\mathcal{I}}_i(nT) \hat{\mathcal{I}}_i(0)\rangle} $, as the tuning parameter. 
The interaction strength is also fixed at ${\lambda=0.7}$ such that ${T\lambda\gg1}$, 
to avoid possible DTC features emerging in the conventional prethermal regime~\cite{Abanin:2017,Else:2017-b,Luitz:2019}. 
Since we are interested in unveiling nonergodic coherent dynamics in a strongly interacting clean model, 
we need the parameter $V$ to be small enough relative to the interaction strength, but remains in the same order of $ \lambda $ such that (\textit{i}) 
``isolated bands''~\cite{Papic:2015} and the resulting nonergodic Floquet dynamics due to finite-size effects~\cite{Seetharam:2018} are not manifested,
(\textit{ii}) for 
accessible system sizes, the model locates far away from its near integrable manifold in the vicinity of exact integrable ${V=0}$ line. 
To firm up the absolute stability of the observed time crystallinity, our main focus is on the generic case $ {V,\Delta\neq0} $ with open boundary condition (OBC), for which the model does not exhibit any explicit microscopic symmetry 
except the global charge conservation, i.e., a physically natural symmetry not requiring fine-tuning.

\section{Time translation symmetry breaking via dynamical scar states}\label{sec:TTSB}
In the first set of calculations we deliver our findings regarding the level spacing ratio ${r_n=\mathrm{min}(\delta_{n+1}/\delta_{n},\delta_{n}/\delta_{n+1})}$ where ${\delta_{n}=E_n-E_{n-1}}$ is the phase gap and $E_n$ denotes n$^\mathrm{th}$ quasi-energy of the Floquet operator. 
The spectrally averaged $r_n$ over symmetry-resolved Hilbert space sectors ($r_\mathrm{ave}$), shown in Fig.~\ref{fig:MLS_OBC_mesh_line}(b), indicates that there exist an apparent phase repulsion in most of the parameter space ${(V,\Delta)}$ explored by $ \mathcal{U}_F $,
as ${0.50\lesssim r_\mathrm{ave}\lesssim 0.53}$ comes close to the Wigner-Dyson value characteristic of quantum many-body chaos~\cite{Atas:2013,Alessio:2014}. 
In Fig.~\ref{fig:MLS_OBC_mesh_line}(c), we also investigated the finite-size behavior of $ {r_\mathrm{ave}(V,\Delta)} $  
at two typical fixed values of $ {\Delta=1} $ 
(top panel), 
and the inversion symmetric  ${\Delta=0}$ line 
(bottom panel). 
For both cases, turning on 
a small value of $ V $
breaks exact integrability of the model
such that for large enough systems sizes, 
the level spacing exhibits a discernible thermal plateau very close to the prediction of circular orthogonal ensemble (COE) distribution, $ {r_{\mathrm{COE}}\approx0.526}$~\cite{Alessio:2014}. 
The near-COE plateaux shorten very slowly in system size
and ultimately end up in a completely chaotic regime, where $ {r_\mathrm{ave}(V)} $ is enclosed by the random matrix values.

The only
exception arising in $ {r_\mathrm{ave}(V, \Delta)} $ is 
a tiny \textit{near integrable} region~\cite{Santos:2010,Modak:2014,Modak:2014_IOP,Luca:2016}(blue area in Fig.~\ref{fig:MLS_OBC_mesh_line}(b)),
that appears in the vicinity of both the exact integrable $ {V=0} $ and $ {\Delta=0} $ lines.
Within this region, $r_\mathrm{ave}$ first fall into a value corresponding to the integrable Poissonian (POI) limit of ${r_\mathrm{POI}\approx0.386}$, yet flows towards thermal value with increasing system size.  
The crossover to the chaos is clearly visible in
Fig.~\ref{fig:MLS_OBC_mesh_line}(d), which displays 
the finite-size behavior of $ r_\mathrm{ave} $ in the integrable looking region, e.g., $ {V\lesssim0.06\ll\lambda} $ at fixed $ {\Delta=0,1} $.
Accordingly, 
the near integrable region is
strongly narrowed by increasing $ N $,
heralding robust thermalization of the model for 
an \textit{arbitrary} small
${V\neq0}$ in the \textit{thermodynamic limit}. 

\begin{figure}[t!]
\centering
\includegraphics[width=1.0\linewidth]{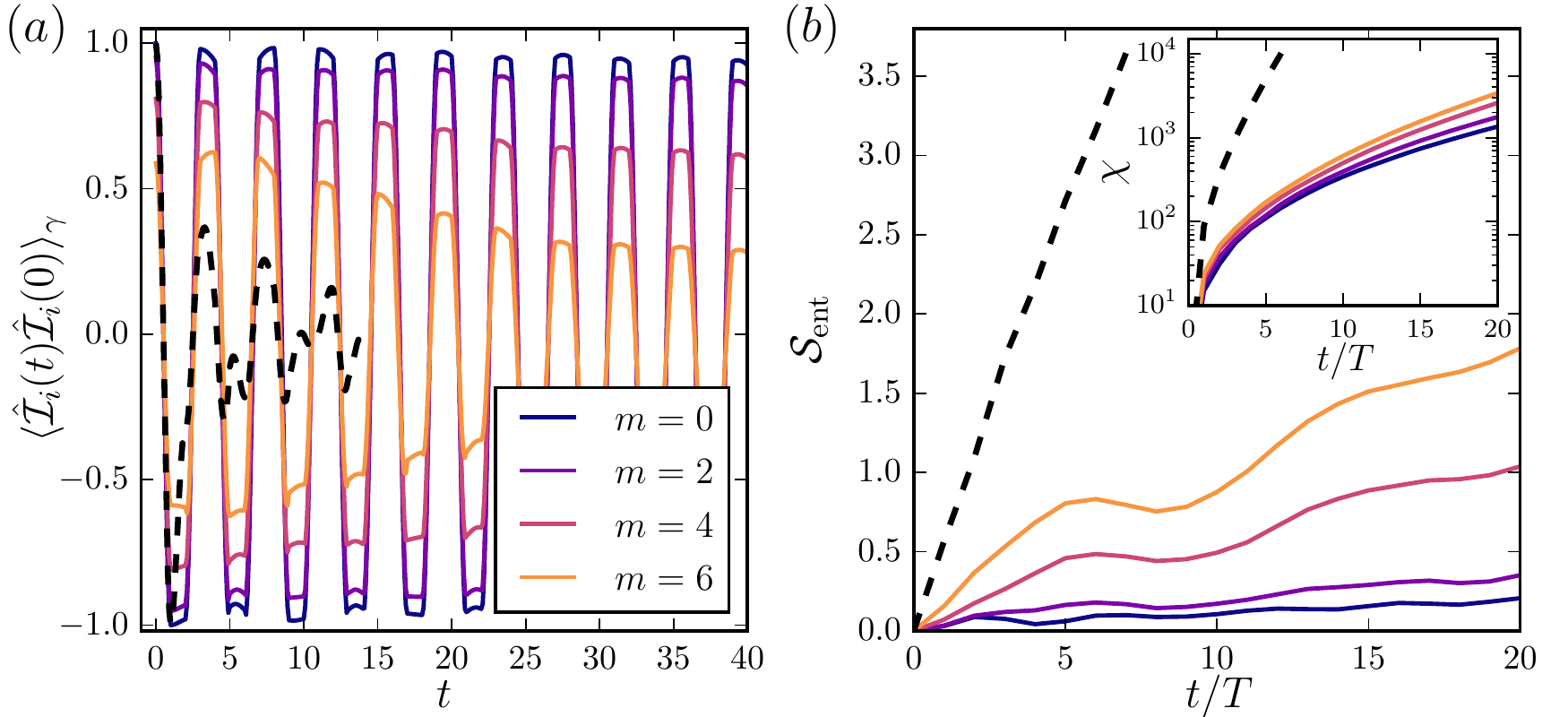}
\caption{(a) Persistent subharmonic oscillations of local imbalance autocorrelator for a typical unit cell, and (b) the entanglement growth for the midpoint bipartition, starting from density-wave product states ${|\psi_\gamma\rangle}$ with ${\gamma = m\pi/40}$, where $ {m=0} $ corresponds to the perfect CDW state. 
In both plots, the solid lines indicate the results in the strongly interacting regime, $ {\lambda=0.7\,\,(=\Delta)} $, ${V=0.1}$.
The dashed lines represent the dynamics of the corresponding quantities for ${m=0}$ in the fully ergodic regime, ${V=\lambda=0.7}$.
These results are provided by iTEBD method
with the maximum iMPS bond dimension ${\chi_{\mathrm{max}}=12000}$ (${4000}$) for ${m=0}$ (${\neq0}$). 	Inset displays the evolution of bond dimension used in the simulation of dynamics. 
}
\label{fig:iTEBD_m} 
\end{figure}

Despite being generically chaotic, the model~\eqref{eq:FM1} can exhibit anomalous discrete TTSB.
To show this, we use the infinite time-evolving block decimation (iTEBD) scheme~\cite{Orus:2008} by the implementation of 
infinite matrix product states (iMPS). 
This method allows to simulate unitary evolution in the infinite volume limit, albeit up to some finite time limited by the maximum bond dimension ${\chi_{\mathrm{max}}\sim 10^3-10^4}$.   
We consider a family of short-range correlated initial conditions of the form 
$ {|\psi_\gamma\rangle=\bigotimes^{N/2}_{i=1} (\cos\gamma\; c^\dagger_{2i} + \sin\gamma\; c^\dagger_{2i+1})|0\rangle} $,  
which are translational invariant up to translations of two lattice spacings. Any nonzero value of $ {\gamma\in(0,\pi/4}) $ introduces an initial-state imperfection with respect to the perfect charge density wave (CDW), i.e., ${|\psi_0\rangle = |\dots 0101 \dots \rangle}$. 
Choosing such translational invariant initial states can immediately rule out the possible realization of nonergodic dynamics that may arise due to ``quasi-MBL'' mechanism~\cite{Papic:2015,Schiulaz:2015,Yao:2016,Yarloo:2018}, 
wherein self-generated disorder is inherited from the inhomogeneity of initial states. 
Figure~\ref{fig:iTEBD_m} represents time series of $\langle\hat{\mathcal{I}}_{i}(t)\hat{\mathcal{I}}_{i}(0)\rangle_\gamma$ in the strong interaction regime  
$ {\lambda=0.7} $, ${V=0.1}$ (solid lines),
which exhibits nonergodic $2T$-oscillations persisting for unusual long times. 
Additionally, the \textit{ballistic} spreading of entanglement entropy~\cite{Alessio:2014,Lazarides:2014}
as well as exponential growth of bond dimension $\chi(t)$~\cite{Prosen:2007}
are both significantly \textit{slower} than those expected to appear in a common thermalizing phase in the opposite extreme limit ${\lambda/V\sim1}$ (dashed lines in Fig.~\ref{fig:iTEBD_m}), where any temporal feature would be entirely absent.

\begin{figure*}[t!]
\centering
\includegraphics[width=1\linewidth]{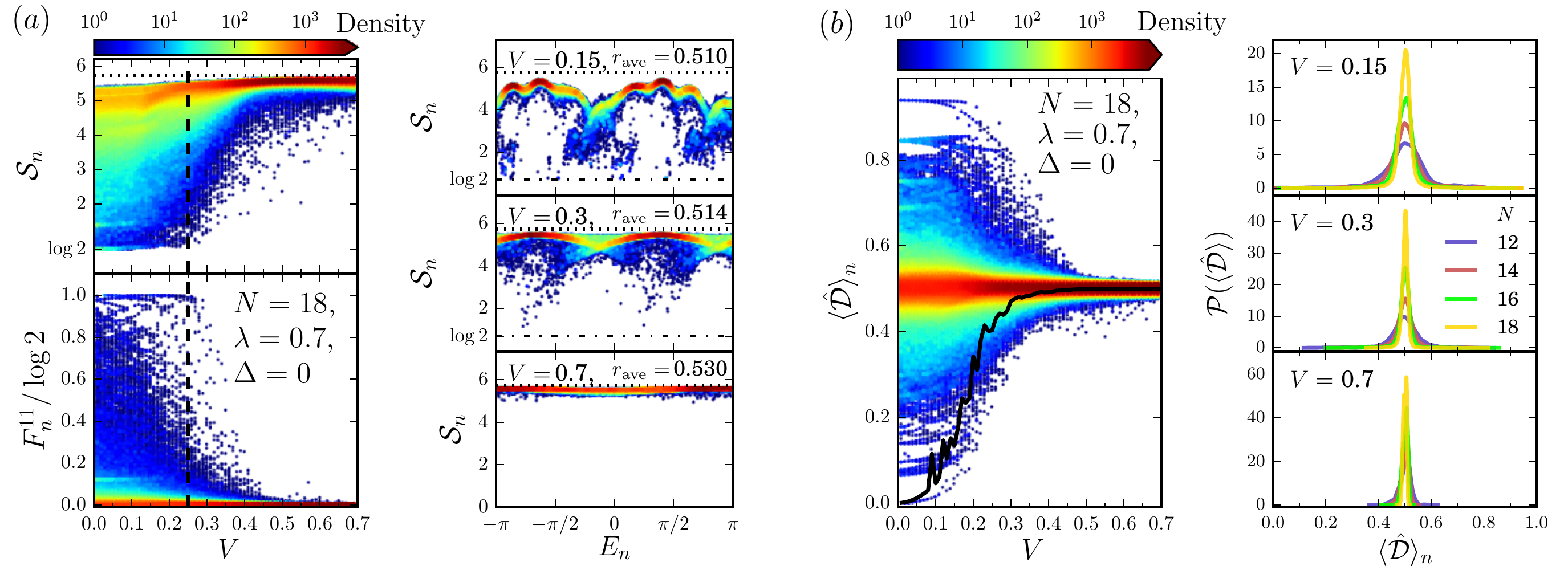}
\caption{(a) Left panel: Scatterplot of the half-cut entanglement entropy (top) and mutual information (bottom) of Floquet eigenstates as a function of $V$ for $ {N = 18} $ and ${ \Delta=0}$. The colorbar denotes the density of points.
For $ {V\lesssim V_{th}} $, denoted by the vertical dashed line, the Floquet eigenstates have a wide range of entanglement values. 
Right panel shows the distribution of half-cut entanglement entropy of eigenstates at three typical fixed values of $ {V=0.15,0.3,0.7} $ as a function of quasi-energy, $ E_n $.
Dotted lines indicate the maximal Page value
for a random pure state~\cite{Page:1993}.
(b) Left panel: Scatterplot of the expectation value of doublon density $\langle\hat{\mathcal{D}}\rangle_n$ as a function of $V$ for $ {N = 18} $, ${ \Delta=0}$. Solid line represents the diagonal ensemble average, $\langle \hat{\mathcal{D}}\rangle_\mathrm{DE}$,  corresponding to the initial CDW state. Right panel: distribution of $\langle\hat{\mathcal{D}}\rangle_n$ at fixed values of $V$, showing strong narrowing with increasing system size. Choosing other values of $ \Delta $ leads to qualitatively similar results. 
}
\label{fig:doublon_Daelta0_OBC} 
\end{figure*}
To provide an initial insight into the nature of such a nonergodic coherent dynamics, we investigate the eigenstate properties of the Floquet unitary in Eq.~\eqref{eq:FM1}. 
For each individual eigenstate $| \Psi_n \rangle$, which is trivially a Floquet steady state, we calculate the half-cut von-Neumann entanglement entropy,
$ {\mathcal{S}_n\equiv\mathcal{S}^{N/2}_n =-\Tr \left( \rho_{N/2} \log \rho_{N/2} \right) }$,
where $ {\rho_{N/2}=\Tr_{N/2} |\Psi_n\rangle \langle \Psi_n|} $ is the reduced density matrix for half system.
We also evaluate the quantum mutual information~\cite{Amico:2008},
defined as 
${F_n^{AB}\coloneqq \mathcal{S}_n^A+\mathcal{S}_n^B-\mathcal{S}_n^{A\cup B}}$, where $ A $ and $ B $ are two spatially separated subsets of the system (in our case, the left- and rightmost 
sites), and $ \mathcal{S}_n^{X}$ 
denotes the entanglement entropy of the reduced density matrix of subset $ {X} $ 
for $ n^{\mathrm{th}} $ Floquet eigenstate. 

In Fig.~\ref{fig:doublon_Daelta0_OBC}(a), we plot the distribution of $ \mathcal{S}_{n} $ and ${F_n^{11}}$ as a function of $ V $ for a system of size ${N=18}$. 
Herein lies the essence of weak ergodicity breaking, giving rise to the anomalous discrete TTSB:  
over all considered ranges of $V$, the majority of Floquet eigenstates look like the entropy-maximizing thermal state near the infinite temperature and are short-range correlated, $ {F_n^{11}\sim0} $.
However, we identify the coexistence of low and high entangled eigenstates over a substantial range of $ {V\lesssim V_{th}\sim 0.25}$, in which the broadening of the entanglement distribution is barely discernible. 
Such a broadening is in sharp contrast to the usual expectations from the Floquet-ETH, but is analogous to that observed in fracturing phenomenon~\cite{Pai:2019,Pai:2019_fractonlocalization,Khemani:2019_shatter,Hudomal:2020,Yang:2019}.
Here, a subset of anomalous nonthermal states, a.k.a., dynamical scars~\cite{Pai:2019}, manifests in the steady-states of Floquet system and can be characterized by their subthermal entanglement.

In particular, for ${V\lesssim V_{th}}$, there are some number of dynamical scars exhibiting anomalous long-range spatial correlation needed for discrete TTSB, i.e., $ {F_n^{11}\sim\log2} $ for $2T$-periodicity~\cite{Else:2016}.
The presence of such a special scar subregion of the Hilbert space can 
underpin spontaneous discrete TTSB in  thermalizing Floquet spectrum,  
when the system is properly initialized in an experimentally accessible, symmetry broken state. 
This leads to the 
formation of 
SDTC dynamics that is 
distinct from traditional MBL-DTCs,    
wherein a finite fraction of nonthermal eigenstates (potentially \textit{all}) 
can feature stable $ \pi $ spin-glass order. 
In the latter case, the $\pi$ spectral pairing structure of entire Floquet spectrum can serve as a practical hallmark of the time crystallinity, which can be quantified using, e.g., $\pi$-translated level spacings~\cite{Keyserlingk:2016}. Obviously, this is not the case when detection of the SDTC is concerned. 

As is clear from Fig.~\ref{fig:doublon_Daelta0_OBC}(a), the dynamical scar states tend to merge with thermal states around $ {V_{th}} $, indicating precursor to the ergodic behavior, and ultimately disappear 
beyond $ {V_{pd}\sim 0.4} $ where 
the onset of Floquet thermalization 
sets in~\cite{Haldar:2018}~(for the reason that will become clear later, we refer to $V_{th}$ and $V_{pd}$ as \textit{thermalization} and \textit{period-doubling} crossover, respectively). 
It is worth noting that in all mentioned ranges of $V$, the spectral statistics is of Wigner-Dyson type, 
and hence cannot distinguish between completely chaotic regime and those containing a vanishing fraction of dynamical scars.

It is also instructive to look at the expectation value of doublon density  ${\hat{\mathcal{D}}=\frac{1}{N/2-1}\sum_i\hat{n}_{i} \hat{n}_{i+1}}$,
measured 
in each individual Floquet eigenstate, $\langle\hat{\mathcal{D}}\rangle_n$. 
As clearly seen in Fig.~\ref{fig:doublon_Daelta0_OBC}(b), 
the main concentration of $\langle\hat{\mathcal{D}}\rangle_n$ is centered around its infinite temperature thermal value at half-filling, 
i.e., ${\langle\hat{\mathcal{D}}\rangle_\infty=1/2}$~\cite{Seetharam:2018}. 
Moreover, the distribution of doublon density,  $\mathcal{P}(\langle\hat{\mathcal{D}}\rangle)$, gradually narrows with increasing system size according to the ETH prediction~\cite{Alessio:2014,Lazarides:2014};
however, this criterion would not generically warrant strong thermalization, 
as it might occur even for an (infinite) integrable system~\cite{Biroli:2010}. 
Here the essential feature is that in the region ${V\lesssim V_{th}}$, there still exists a strong support within $\mathcal{P}(\langle\hat{\mathcal{D}}\rangle)$ (see the right panel of Fig.~\ref{fig:doublon_Daelta0_OBC}(b)), which stems from the rare existence of athermal eigenstates  with the values of $\langle\hat{\mathcal{D}}\rangle_n$ strikingly different from  $\langle\hat{\mathcal{D}}\rangle_\infty$. 
Such Floquet outlier states may have a considerable overlap with the initial conditions $ \ket{\psi_\gamma} $, and hence significant weight for the corresponding steady-state described by the diagonal ensemble average, $\langle \hat{\mathcal{D}}\rangle_\mathrm{DE}$~\cite{Biroli:2010,Haldar:2018}. 
This fact is evinced from
the deviation of $\langle \hat{\mathcal{D}}\rangle_\mathrm{DE}$ (black solid line in Fig.~\ref{fig:doublon_Daelta0_OBC}(b)) from  $\langle\hat{\mathcal{D}}\rangle_\infty$, 
that reflects the failure of strong ETH, and possible slow relaxation of generic local observables.

\section{Persistence, Crossover and Rigidity}\label{sec:PRC} 
The observed anomalous discrete TTSB leads us to directly examine time-crystalline signature and its fundamentally distinct origin in the presence of quantum many-body chaos.

\begin{figure}[t!]
\centering
\includegraphics[width=1\linewidth]{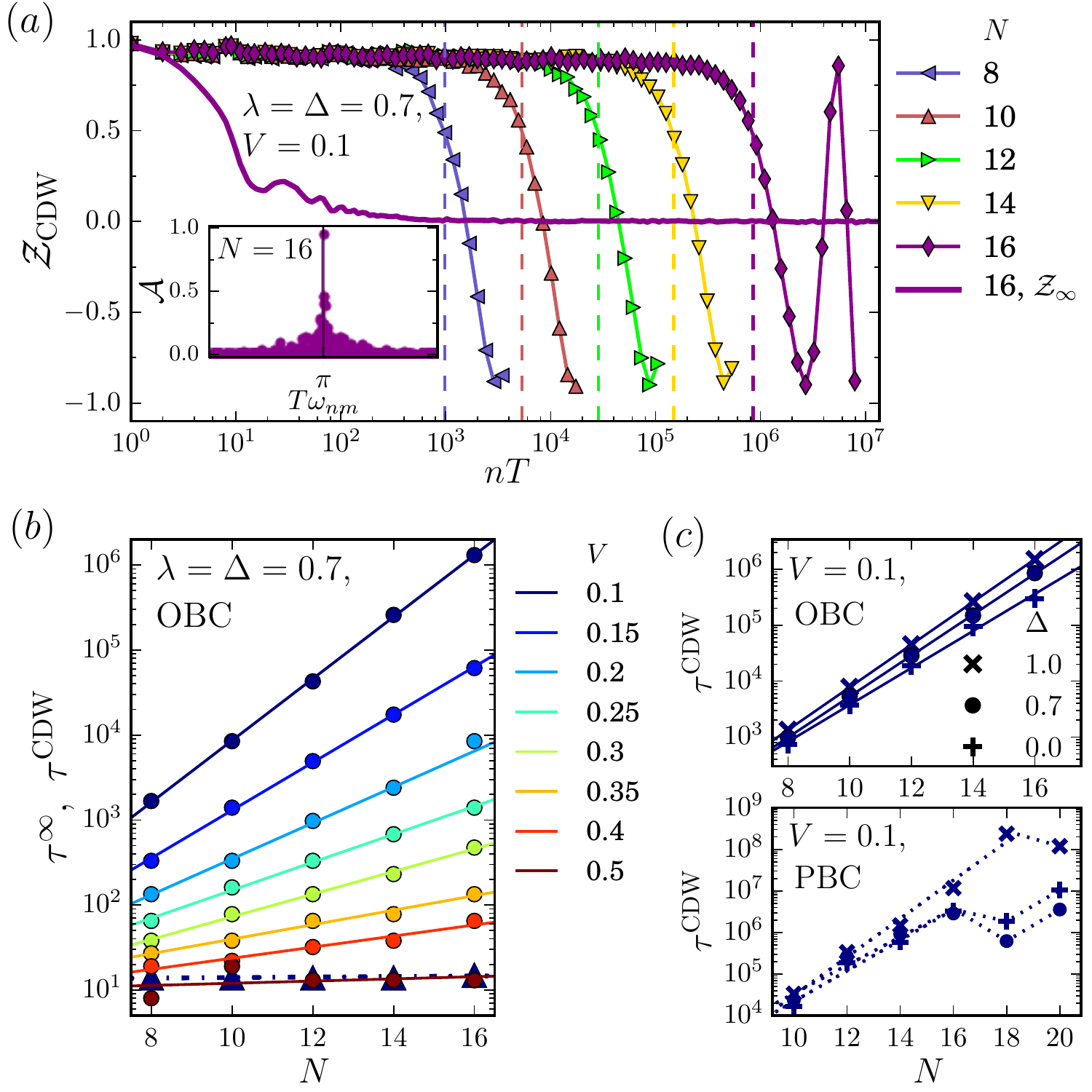}
\caption{(a) The stroboscopic dynamics of time crystal order parameter starting from a perfect initial CDW state at ${V=0.1}$. 
Inset displays the spectral weight peaked at ${T\omega_{nm}\sim\pi} $. In contrast to $\mathcal{Z}_\infty$ (solid line), obtained from Eq.~\eqref{eq:infZ}, the decay of $\mathcal{Z}_\mathrm{CDW}$ exhibits a long-lived DTC plateau with an exponentially diverging time scale set by ${|T\omega_{p}-\pi|^{-1}}$ (dashed lines). 
(b) The scaling behavior of 
the crystalline melting times (circles) for various values of $ V $.
Triangles denote $\tau^{\infty}$s extracted from $\mathcal{Z}_\infty$ for ${V=0.1}$. (c) Upper panel: The qualitative independence of the melting times on the ionic potential. Bottom panel demonstrates the existence of a 
characteristic length scale $\ell_c$, at which ${\tau^{\mathrm{CDW}}}$ stops its initial exponential growth.
}
\label{fig:OBC_dynamics} 
\end{figure}

\emph{\textbf{Persistence}---.} To settle the dynamical fingerprint of the scar states, we first investigate the persistence of subharmonic response as well as its initial-state dependence. To this end, we evaluate dynamics of the stroboscopic-time staggered total density imbalance, $ {\hat{\mathcal{I}}_{tot}=2/N\sum_i \hat{\mathcal{I}}_{i} }$, evolving from an initial CDW state, 
\begin{align} \label{eq:CDWZ}
{\mathcal{Z}_{\mathrm{CDW}}(nT)\equiv\langle \psi_0| (-1)^n \hat{\mathcal{I}}_{tot}(nT) \hat{\mathcal{I}}_{tot}(0) |\psi_0\rangle},
\end{align}
as a measure of the time crystallinity~\cite{Else:2016}. 
To track the manifestation of DTC order
in an initial-state \textit{independent} manner, it is convenient to consider the normalized Hilbert-Schmidt distance~\cite{Abanin:2017},
\begin{align} \label{eq:}
\varrho_{\mathrm{HS}}(nT)
\equiv \small{\dfrac{\vert\vert \hat{\mathcal{I}}_{tot}(nT) - (-1)^n\hat{\mathcal{I}}_{tot}(0)\vert\vert_{\infty}^2}{2\vert\vert \hat{\mathcal{I}}_{tot} \vert\vert_{\infty}^2} }
=1- \mathcal{Z}_\infty(nT),
\end{align}
where,
\begin{align} \label{eq:infZ}
 { \mathcal{Z}_\infty(nT)=\dfrac{1}{\vert\vert \hat{\mathcal{I}}_{tot}\vert\vert_{\infty}^{2}} \Tr((-1)^n \hat{\mathcal{I}}_{tot}(nT) \hat{\mathcal{I}}_{tot}(0))}, 
\end{align}
and $ \vert\vert\,\cdots\,\vert\vert_{\infty} $ denotes the Hilbert-Schmidt operator norm.
Obviously, $ {\mathcal{Z}_\infty(nT)\neq0} $ (or equivalently $ {\varrho_{\mathrm{HS}}(nT)\neq1} $) when the Floquet dynamics exhibits DTC order. 
Autocorrelations of this type have also been used to identify the survival of 
the MBL-driven and prethermal $ \mathrm{U}(1) $ DTCs at infinite temperature~\cite{Yao:2017,Luitz:2019}. 
The results for a typical small value of ${V=0.1}$, shown in Fig.~\ref{fig:OBC_dynamics}(a), signify that ${\mathcal{Z}_\infty(nT)}$ decays rapidly to zero. The same result also holds for the autocorrelators of local $ {\hat{\mathcal{I}}_i} $ operators.
So, any time crystal feature gets lost for this case. 
By contrast, $\mathcal{Z}_\mathrm{CDW}(nT)$ first drops to a smaller nonvanishing value followed by a long-lived plateau which eventually terminates by some finite-size revivals 
at the late times. 
This clearly adds to a strong initial-state dependence in the time crystallinity observed in the presence of many-body chaos.

To give system-size dependence of the crystalline melting time,
followed by Refs.~\onlinecite{Keyserlingk:2016,Huang:2018}, we
calculate the spectral weight for temporal correlator of imbalance operator, i.e.,
${\mathcal{A}(\omega_{nm})=|\langle \Psi_n| \hat{\mathcal{I}}_{tot} | \Psi_m \rangle|^2}$ where ${\omega_{nm}=E_n - E_m}$.
This quantity is sharply peaked close to ${\omega_{p}\sim\pi/T}$ (see the inset of Fig.~\ref{fig:OBC_dynamics}(a)) and the plateaus in $\mathcal{Z}_\mathrm{CDW}$ fall off at times 
inversely 
proportional to the perfectness of this $ \pi $-pairing, i.e.,  ${|T\omega_{p}-\pi|^{-1}\sim e^{\mathcal{O}(N)}}$. 
As shown in Fig.~\ref{fig:OBC_dynamics}(b), for ${V\lesssim V_{th}}$ the melting times experience a strong system-size dependence compared to the $N$-independent ${\tau^{\infty}}$s extracted from $ \mathcal{Z}_\infty $. In this regime, the scaling behavior of ${\tau^{\mathrm{CDW}}}$ remains almost independent of $\Delta$ (upper panel of Fig.~\ref{fig:OBC_dynamics}(c)), and hence is relatively insensitive to  microscopic (explicit) symmetries of the model.  
By further increasing $ V $, the system-size dependence of ${\tau^{\mathrm{CDW}}}$ becomes weaker and ultimately diminishes for ${V\gtrsim V_{pd}}$, consistent with the ETH expectations. 

From the first sight, the observed DTC response shows exactly the same diagnostics as those of the parent nonergodic model ${\mathcal{U}_{F}^{int}(\varepsilon,\lambda,\Delta)}$, 
whose level spacing in its DTC regime does remain close to Poisson statistics as $ {N\to\infty} $~\cite{Huang:2018}.
In particular, the subharmonic oscillations appear to exist for an infinitely long time.
However, going to larger system sizes by taking into account periodic boundary condition (PBC) unveils that the exponential growth of ${\tau^{\mathrm{CDW}}}$ 
persists only up to a finite time scale $ \tau_c $, associated with a characteristic length scale $\ell_c\equiv\ell({V,\Delta})$ (see bottom panel of Fig.~\ref{fig:OBC_dynamics}(c)).
For $ {N<\ell_c} $, 
the $ \pi $-pairing looks perfect and 
the period-doubled response is preserved over exponentially long time scale $ \tau_c $. 
For ${N>\ell_c}$, the melting time gradually stops its initial growth, while still being exponentially large compared to $ \tau^\infty $. 
However, given the bound on the crystalline melting time,
we cannot rule out the possibility of eventual thermalization 
which might occur
at much longer time/length scales than are accessible to our numerics.
This resembles generic (time-independent)
scarred Hamiltonians, 
in which a similar thermalization length scale controls  late-time dynamics and the  survival of quantum scars; see, e.g., Refs.~\onlinecite{Khemani:2019,Lin:2019} for the evidence of $ \ell_c $ in the so-called PXP model in the absence of fine-tunning.

Our present investigations first suggest that
the homogeneous DTC 
in the generically chaotic model~\eqref{eq:FM1},
should in principle be exponentially but \textit{not} necessary infinitely long-lived; 
the fact that marks the \textit{partial} persistence of the SDTC.
This 
is in sharp contrast to 
the traditional MBL-DTCs that are absolutely stable in the limits $ {t\to\infty} $ and $ {N\to\infty} $, and fulfill the strict definition of time crystals~\cite{Keyserlingk:2016}.
Second, our results rule out the realization of a true time crystal in ${\mathcal{U}_{F}^{int}(\varepsilon,\lambda,\Delta)}$, as its lifetime does not strictly extend to infinity upon adding generic $ V $-perturbation (even at ${\varepsilon=0} $). 
Moreover, as previously mentioned in Figs.~\ref{fig:MLS_OBC_mesh_line}(b-d), the signatures of integrability in the 
spectrum of  ${\mathcal{U}_{F}^{int}}$ will be completely lifted by an arbitrary small $ {V\neq0} $ 
in the thermodynamic limit.  
Therefore, the Floquet integrability emerging in the bulk spectrum of ${\mathcal{U}_{F}^{int}}$~\cite{Huang:2018} cannot generally provide a stable protecting mechanism for realizing a true DTC phase in generic clean systems.

\emph{\textbf{Thermalization crossover}---.} To specify the boundary between coherent and thermal regimes, 
from \textit{dynamics}, we apply a method from machine learning based on the ``confusion'' scheme~\cite{vanNieuwenburg:2017}, yet with employing recurrent neural networks (RNN) architecture instead of their more common nonrecurrent variants, namely  feed-forward networks~\cite{vanNieuwenburg:2017,Venderley:2018,Kharkov:2020}. 

The RNNs are designed for processing sequential data with a kind of \textit{memory} (see e.g., Ref.~\onlinecite{vanNieuwenburg:2018}). However, as a supervised method, it requires training on correctly labeled input-output pairs in the extremities of the phase space. Thus it is not directly applicable for the problem where labeling is not known beforehand, especially, from the perspective of \textit{finite-size} and \textit{finite-time} data.  
On the other hand, the heart of the semisupervised confusion algorithm is based on the purposefully mislabeling the input data through proposing dummy critical point $V_d$, and then evaluating the total performance of a trained network with respect to the proposed $V_d$s. It is expected that the network performance takes a characteristic universal W-shape as a function of $V_d$, whose middle peak at $V'_d$ implies the correct labeling associated with true critical point~\cite{vanNieuwenburg:2017}; as it would be easiest for the network to classify data for this choice of separation. 
Tending to do so, the confusion scheme through finding the majority label for the underlying (hidden) structure of dynamics, can be utilized to help RNN in the task of detecting SDTC to Floquet-ETH crossover using a prior \textit{unknown} labels.
\begin{figure}[t!]
\centering
\includegraphics[width=1\linewidth]{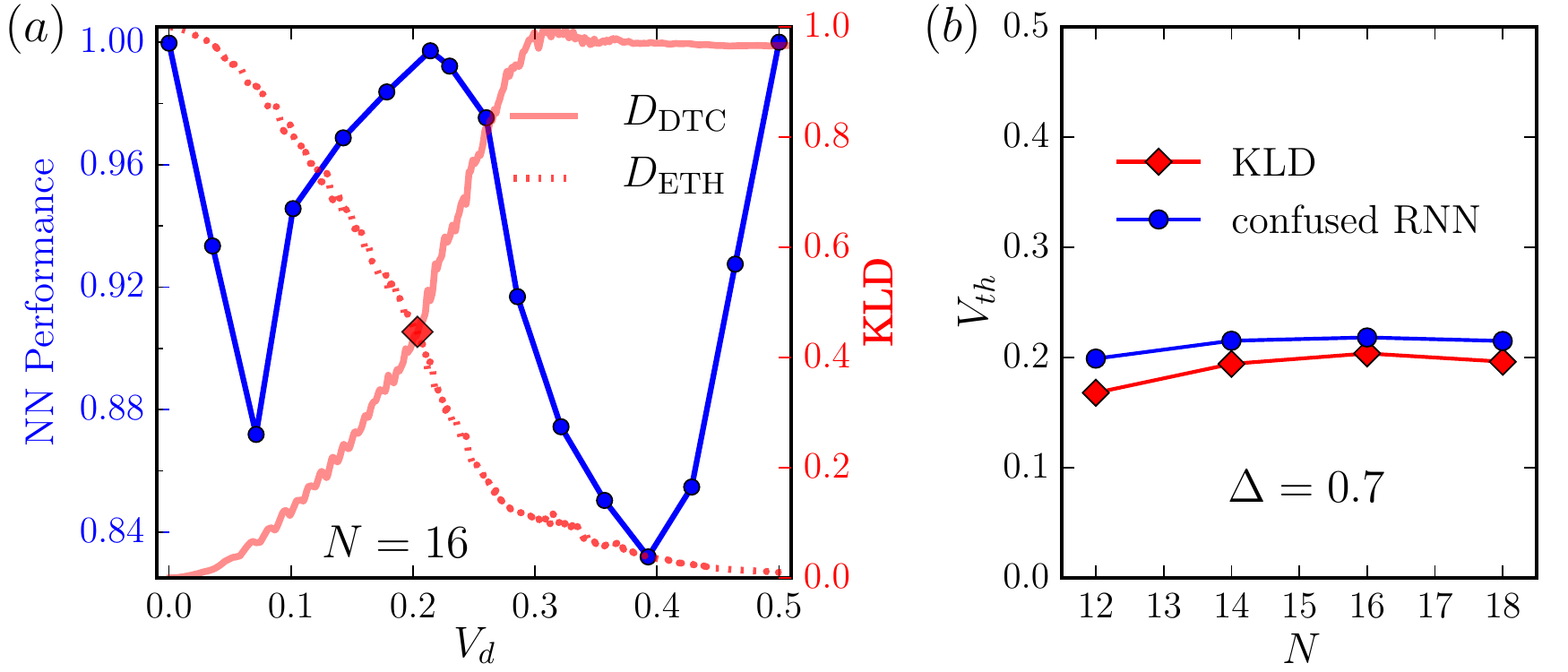}
\caption{(a) Universal W-like NN performance curves (left axis) in the confused RNN for the model~\eqref{eq:FM1} with ${N=16}$, ${\lambda=\Delta=0.7}$ and a fixed set of learning parameters: ${l_2 = 0.01} $,  $ {\alpha = 10^{-5}} $, dropout $ 0.2 $, batchsize of $ 100 $ and $ 400 $ training epochs. The middle peak pinpoints the exact value of the transition at ${V_{th} \approx 0.22}$ that coincides with the prediction of the KLD calculation in Eq.~\eqref{eq:KLD} (right axis). (b) System-size dependence of $V_{th}$ predicted from the machine learning (circles) and KLD analysis (squares). Repeating this procedure for different values of $ \Delta $, leads to the phase diagram shown in Fig.~\ref{fig:MLS_OBC_mesh_line}(b). 
}
\label{fig:RNN_confusion} 
\end{figure}

We proceed with training a RNN on the stroboscopic time-series of ${C_i(nT)\equiv\langle\hat{\mathcal{I}}_{i}(nT)\hat{\mathcal{I}}_{i}(0)\rangle}$ with ${i = 1, \dots ,N/2}$, evaluated during the first ${n_\mathrm{max}=1000}$ periods and sampled at five  equally spaced points. 
Thus the input to our networks is of shape ${(N/2, n_\mathrm{max}/5)}$.  We choose a single hidden layer network only with $16$ long short-term memory
units~\cite{Hochreiter:1997}, for fixed batchsize $100$ and $400$ epochs. In each epoch all training data lie inside the proposed range of~${V_d\in[0, 0.5]}$.
The actual training of the network is done by $ 8000 $ samples, the learning rate $ {\alpha = 10^{-5}} $, a dropout rate~\cite{Nitish:2014} of $0.2$, and minimizing the cross-entropy using Adam optimizer with weight decay ($l_2$~regularization) of $0.01$, followed by a final softmax layer of size $ 2 $, corresponding to the ergodic and SDTC classes we are distinguishing. 

Figure~\ref{fig:RNN_confusion}(a) reveals the W-like NN performance curve that puts ${V'_d\equiv V_{th}\sim 0.22}$ as true thermalization crossover, which is consistent with the previously estimated value extracted from the structure of \textit{static} data, e.g., the distribution of eigenstates entanglement entropy or mutual information shown in Fig.~\ref{fig:doublon_Daelta0_OBC}(a).
Additionally, the position of central peak remains merely intact with system size (Fig.~\ref{fig:RNN_confusion}(b)), and is almost independent of the ionic potential~\footnote{The location of the middle peak is also robust to small variations in the learning parameters~\cite{vanNieuwenburg:2017}.} (see the phase diagram of Fig.~\ref{fig:MLS_OBC_mesh_line}(b)). 
One can further verify this crossover through the Kullback-Leibler divergence (KLD)~\cite{Kullback:1951}, 
\begin{align} \label{eq:KLD}
{D_{\mathrm{ref}} (V)=\sum _\omega \mathcal{F}_V(\omega)\, \log \,(\mathcal{F}_V(\omega)/\mathcal{F}_{\mathrm{ref}}(\omega)), }
\end{align} 
which measures the distance between the normalized Fourier spectrum of $ \langle\hat{\mathcal{I}}_{tot}(nT)\rangle $ at a fixed $ V $, denoted by $\mathcal{F}_V(\omega) $, and
a reference signal corresponding to either a perfect DTC or a completely chaotic response, denoted by $\mathcal{F}_{\mathrm{DTC}}$ and $\mathcal{F}_{\mathrm{ETH}}$, respectively. At the true critical point $ V_{c} $, one expects $ \mathcal{F}_{V_{c}}(\omega) $ to be equidistant from both $\mathcal{F}_{\mathrm{DTC}} $ and $\mathcal{F}_{\mathrm{ETH}} $, and thereby ${D_{\mathrm{DTC}} (V_{c})=D_{\mathrm{ETH}} (V_{c})} $~\cite{Kharkov:2020}. 
As is clear from Fig.~\ref{fig:RNN_confusion}, this condition is fulfilled for $ {V_{c}\sim0.20~(\approx V_{th})}$, and the critical points extracted in this manner coincide very well with the predictions of the confused RNN.
\begin{figure}[t!]
\centering
\includegraphics[width=1.0\linewidth]{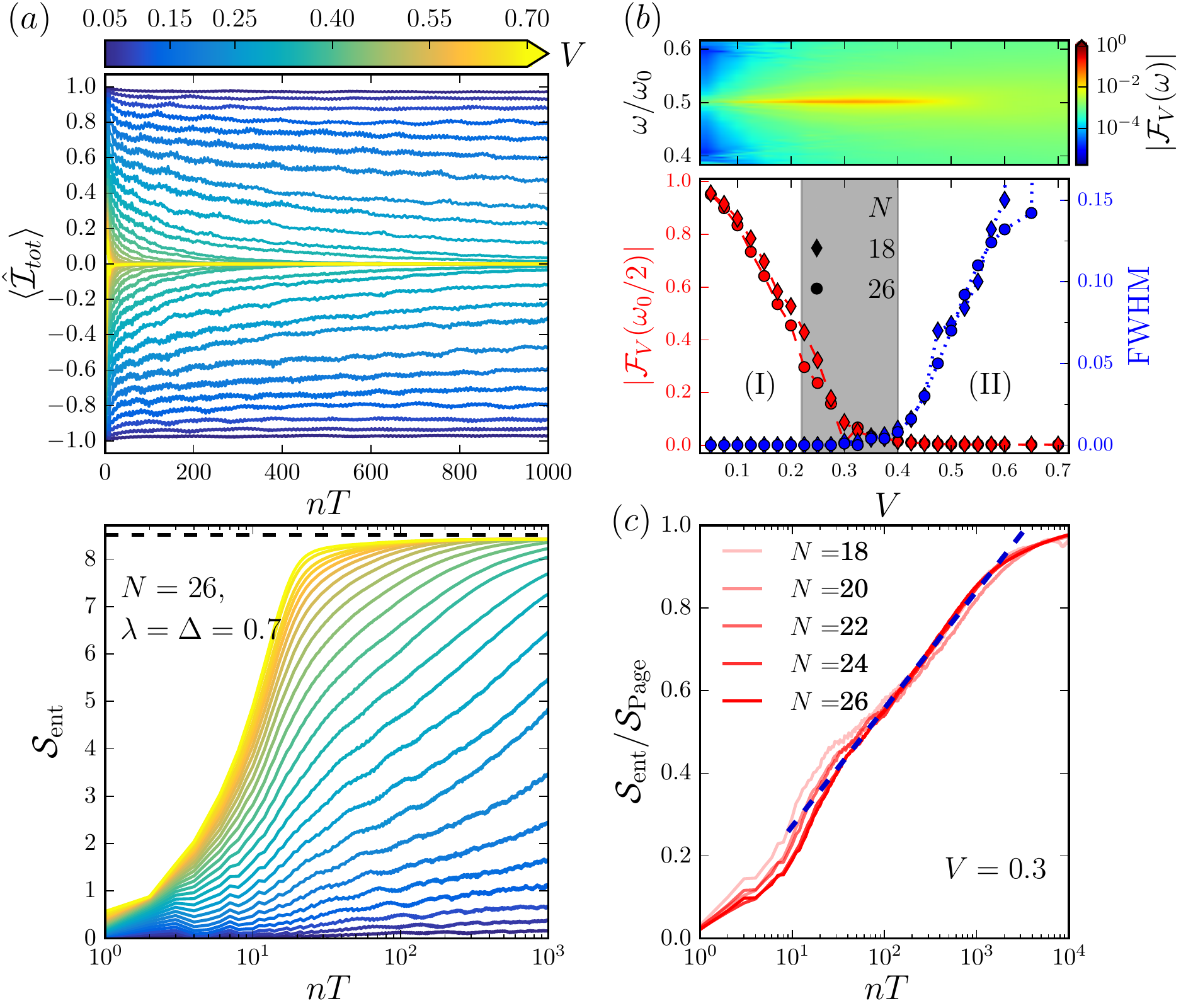}
\caption{(a) The stroboscopic evolution of total density imbalance (upper panel) and entanglement entropy (bottom panel) for various values of ${V\leq\lambda}$ with $ {\lambda=\Delta=0.7} $ and system of size $ {N=26} $. Dashed line indicates the thermal Page value, $\mathcal{S}_{\mathrm{Page}}$. 
(b) The color map represents the normalized power spectrum of       ${\langle\hat{\mathcal{I}}_{tot}(nT)\rangle}$ shown in panel (a). Bottom panel shows the magnitude of the
${\omega_0/2}$
peak (left axis), and the corresponding FWHM (right axis) as a function of $V$ for different system sizes. (c) Bounded logarithmic growth of entanglement entropy (normalized by the Page value) for a typical value of ${V=0.3}$ at the ergodic side of thermalization crossover, $ {V_{th}\approx0.22} $. 
}
\label{fig:cheby_dynamics} 
\end{figure}

\emph{\textbf{Rigidity}---.} Here we investigate the robustness of the SDTC dynamics 
in the finite but thermodynamically large system of size ${N=26}$. Using numerically exact Krylov space based algorithm, 
we evaluate the stroboscopic dynamics of the total
density imbalance ${\langle\hat{\mathcal{I}}_{tot}(nT)\rangle}$, and half-cut entanglement entropy $\mathcal{S}_{\mathrm{ent}}(nT)$ as function of $ V $.  
The results depicted in Fig.~\ref{fig:cheby_dynamics} suggest three distinct dynamical regimes characterized by $V_{th}$ and $V_{pd}$:  
for small imperfection strength ${V\lesssim V_{th}}$ (region~I in Fig.~\ref{fig:cheby_dynamics}(b)), ${\langle\hat{\mathcal{I}}_{tot}(nT)\rangle}$ displays robust $ 2T $-oscillations locked at half of the driving frequency ${\omega_0/2}$. Moreover, the amplitude of the peak, and hence of the oscillations, in the power spectrum  ${|\mathcal{F}_{V}(\omega_0/2)|}$ is apparently large.

In the intermediate regime,  ${V_{th}\lesssim V\lesssim V_{pd}}$, at the ergodic side of thermalization crossover (shadow region of Fig.~\ref{fig:cheby_dynamics}(b)), 
the broadening of the full width at half maximum (FWHM) still remains negligible, heralding the persistence of period-doubled dynamics. 
Nonetheless, the system displays a precursor to thermalizing dynamics: $\mathcal{S}_{\mathrm{ent}}(nT)$ typifies a logarithmic slow growth leading up to an inevitable thermalization at the late times (see Fig.~\ref{fig:cheby_dynamics}(c)).
However, the time interval over which this logarithmic growth happens does not extend with system size, conveying a \textit{bounded} rather than unbounded slow heating characteristic of 
disordered MBL systems~\cite{Bardarson:2012,Vosk:2013,Serbyn:2013_1}. 

Lastly, for $ {V\gtrsim V_{pd}} $ in region~II, the FWHM becomes much more pronounced and any temporal feature would entirely disappear. The entanglement dynamics also changes its own behavior from an extremely slow growth in region~I, to a fast one in region~II where $\mathcal{S}_{\mathrm{ent}}$ quickly approaches the maximal Page value within the time scales accessible by our numeric. The latter feature also verifies that the system size considered here is thermodynamically large enough to warrant the immunity of our results against finite-size effects.
\begin{figure}[t!]
\centering
\includegraphics[width=1.0\linewidth]{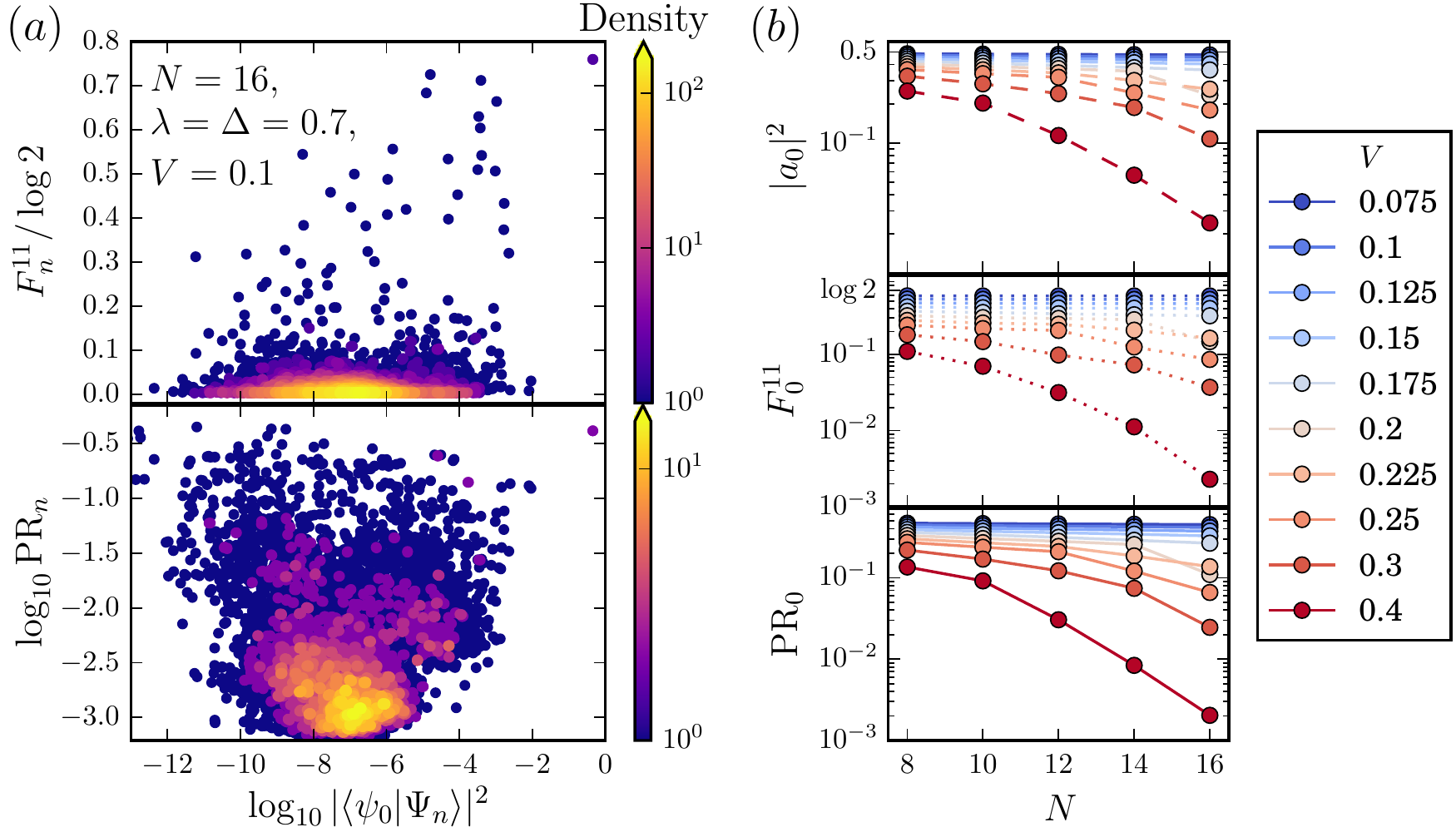}
\caption{(a) Histogram of the overlap of CDW product state with Floquet eigenstates plotted against their mutual information (top) and second participation ratio (bottom) in the deep SDTC regime, $ {\lambda=\Delta=0.7} $, $ {V=0.1} $. (b) From top to bottom: Finite-size scaling of the largest overlap, its respective mutual information and participation ratio for various drive imperfection strengths. We note that the scaling behavior of $ {F_n^{ll}} $ remains qualitatively the same for $ {l>1} $.
}
\label{fig:OBC_N16_V01_Delta07} 
\end{figure}

To firm up the observed period-doubling effect, and its stability, as a direct dynamical manifestation of scar states, we shed light on the structure of Floquet spectrum, when one arranges Floquet eigenstates according to their overlap with CDW state ${|a_n|^2= | \langle \psi_0| \Psi_n \rangle|^2}$, together with their mutual information ${F_n^{11}}$ and second participation ratio ${\mathrm{PR}_n= \sum_\alpha | \langle \alpha| \Psi_n \rangle|^4}$.
The results shown in Fig.~\ref{fig:OBC_N16_V01_Delta07}(a) signify that even deep in the SDTC regime, 
the dominant eigenstates are short-range correlated, delocalized states with an exponentially small ${F_n^{11}}$ and $ \mathrm{PR}_n $, which cannot exhibit symmetry breaking.
However, they are irrelevant with respect to the initial condition,
and hence cannot impede spontaneous discrete TTSB in a striking sense. 
Instead, the dynamics is dominated by special outlier states, which are localized on some subsets of thermalizing Floquet spectrum and display nontrivial spatial correlations.

Figure~\ref{fig:OBC_N16_V01_Delta07}(b) demonstrates the scaling of the largest overlap ${|a_0|^2={\rm{max}}\{|a_n|^2 \}} $, corresponding to the eigenstate with the most considerable weight in dynamics, as well as its respective 
$F^{11}_0$ and 
$\mathrm{PR}_0$. Within the SDTC regime, we find the values  ${|a_0|^2\sim\mathcal{O}(1/2)}$, $ {F^{11}_0\sim\log2} $ and ${\mathrm{PR}_0\sim\mathcal{O}(1)}$, whose scaling remains fairly constant with system size and only exhibits slow decay upon approaching $V_{th}$. Beyond this regime, they do appear to be decreasing exponentially with $ N $; the behavior which becomes much more pronounced as system size increases. 
These results give a clear illustration of the intimate connection between the rigidity of the SDTC dynamics and the robustness of dynamical scars. 

\subsection{Coherent thermalizing dynamics: analogues to Floquet supersymmetry}\label{sec:AFSUSY}
Here we investigate how the presence of dynamical scars affects dynamics of certain observables, \textit{irrespective} of specific choice of initial condition. 
We explore the behavior of the autocorrelation function, ${C^{\infty}_{tot}(nT)\equiv\langle\hat{\mathcal{I}}_{tot}(nT)\hat{\mathcal{I}}_{tot}(0)\rangle_\infty }$, which takes the same form as Eq.~\eqref{eq:infZ} without the factor $ {(-1)^n}
$.
Here our main focus is on the SDTC regime, 
where the scar states has a tangible effect on the manifestation of discrete TTSB, once the system evolves from a simple product CDW state. 
In this regime, the Floquet operator can be restricted to the space spanned by the athermal dynamical scar states, as well as its complement subspace containing otherwise ergodic eigenstates, i.e., ${\mathcal{U}_F = \mathcal{U}_F \mathcal{P}_S+ \mathcal{U}_F (1-\mathcal{P}_S)}$, where $\mathcal{P}_S$ is the projection onto the scarred subspace. 

\begin{figure}[t!]
\centering
\includegraphics[width=1\linewidth]{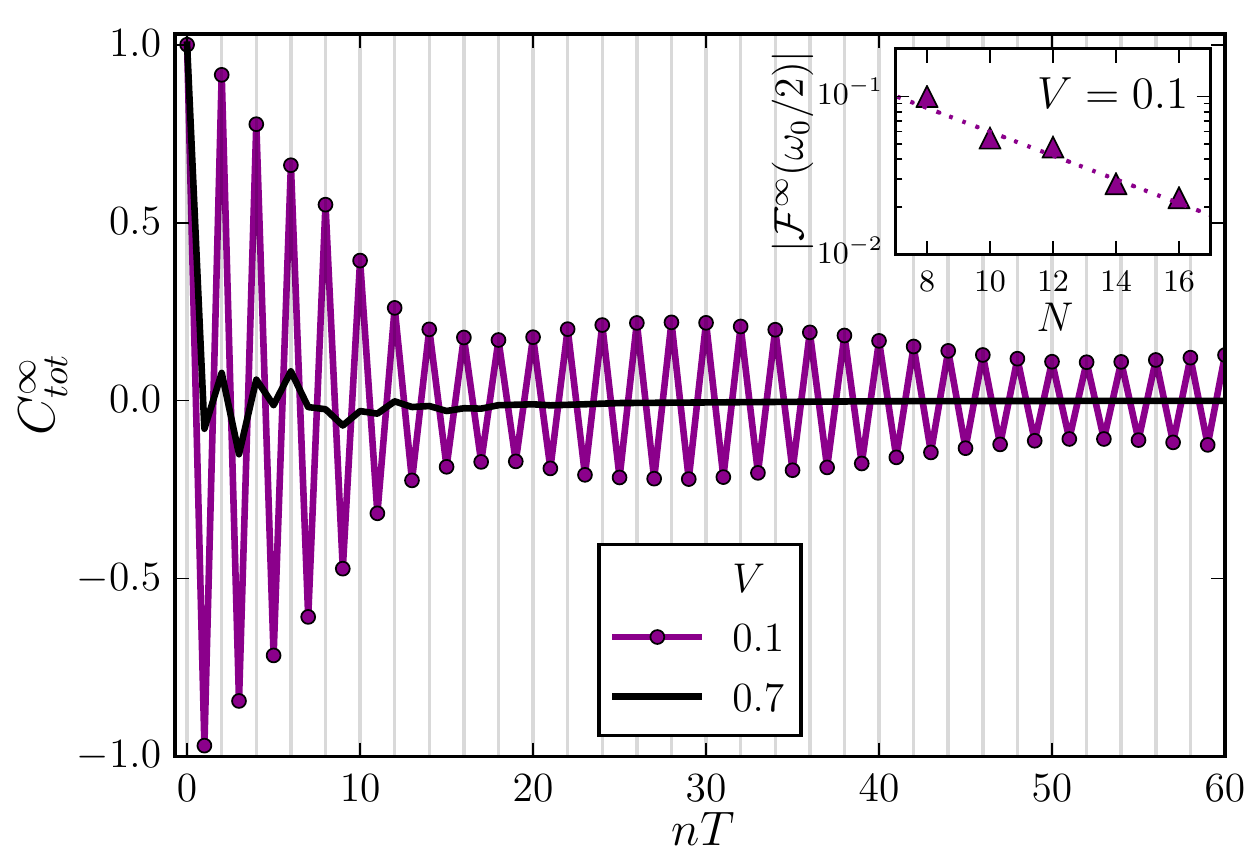}
\caption{The representative time-series of $C^{\infty}_{tot}(nT) $ for system size $ {N=16} $, ${\lambda=\Delta=0.7}$, and ${V=0.1}$ ($ {V=0.7} $) corresponds to the SDTC (thermal) regime. Inset shows the exponential suppression of the magnitude of the 
${\omega_0/2}$ 
peak in the normalized  power spectrum during the first $ 1000 $ periods for ${V=0.1}$.
}
\label{fig:m_dynamics} 
\end{figure}

Figure~\ref{fig:m_dynamics} displays a number of representative time-traces for $C^{\infty}_{tot}(nT)$ in the extremities of the phase space. For ${V=0.7}$  where the dynamics is controlled by a set of thermal states, there are no persistent oscillations.
However, for $ {V=0.1} $ corresponds to the SDTC regime, $C^{\infty}_{tot}(nT) $ exhibits short-time period-two oscillations around its infinite-temperature value. Strikingly, the magnitude of the ${\pi/T}$ peak in the power spectrum,  ${|\mathcal{F}^\infty(\omega_0/2)|}$,
is exponentially decaying with increasing system size (see the inset). 
Hence, the subharmonic oscillation of $C^{\infty}_{tot}(nT) $ can 
persist only at short times and die off exponentially fast in system size, as opposed to a true time crystal.
Here, the contribution of ${\mathcal{U}_F (1-\mathcal{P}_S)}$ tends to drive system towards eventual thermalization, yet with an oscillatory response attributed to the component ${\mathcal{U}_F \mathcal{P}_S}$. The finite-size suppression of oscillations can also be understood through exponential diminution of the number of special scar states with respect to the entire Hilbert space. 
This nontrivial thermalizing dynamics---with robust $2T$-oscillation of certain observables---is distinct from those prescribed by the conventional Floquet-ETH 
with no definite frequency.
\begin{figure*}[t!]
\centering
\includegraphics[width=1\linewidth]{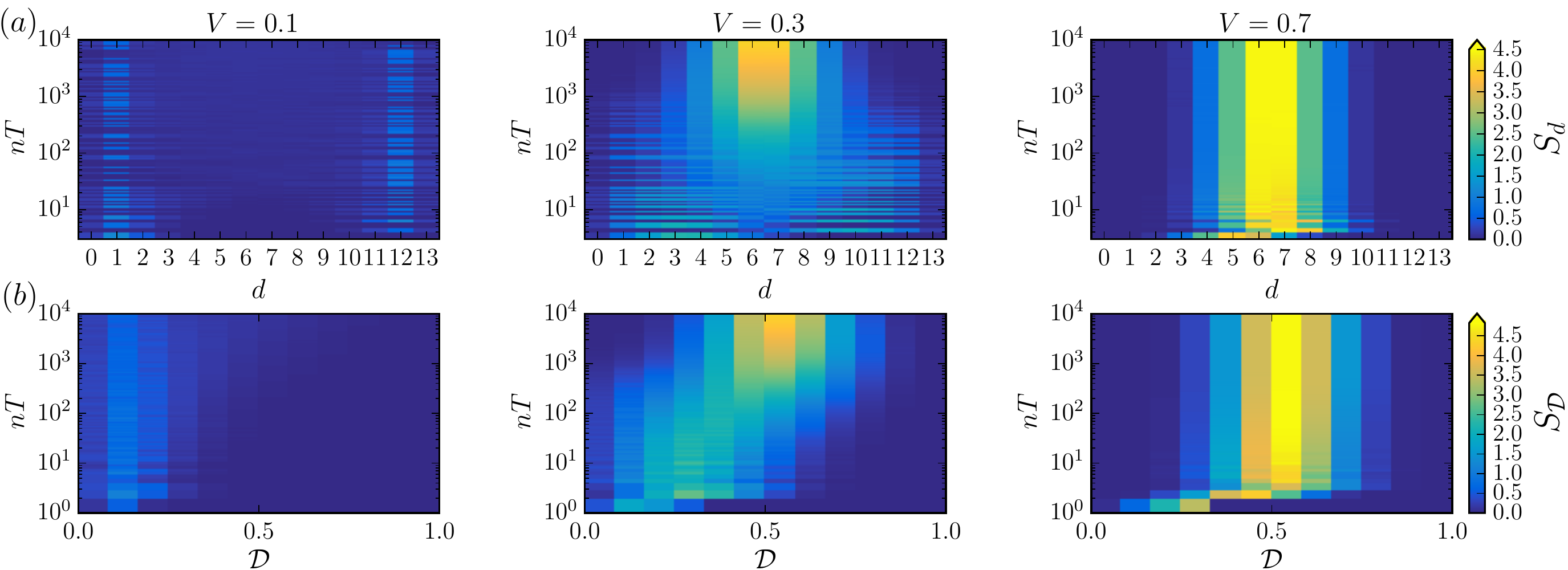}
\caption{The participation entropy of the wavefunction $ |\psi(nT)\rangle $ for $ {\lambda=\Delta=0.7} $ and $ {N=26} $, resolved into the sectors with a fixed Hamming distance $ d $ from the initial CDW state (a), and fixed doublon density $ \mathcal{D} $ (b). From these results, three qualitative distinct regimes are evinced: small $ {V\lesssim V_{th}} $ consistent with the SDTC regime, wherein the wavefunction remains localized in the vicinity of $ d_{min} $ and $ d_{max} $, and cannot evolve out to other intervening sectors; large $ {V\gtrsim V_{pd}} $, in which the rapid expansion of the initial wavefunction indicates a fast approach to infinite temperature; and a crossover regime at intermediate $ {V} $, where a precursor to ergodic dynamics forms.
}
\label{fig:D_cons} 
\end{figure*}

Such coherent approach to thermal equilibrium emerging in a \textit{finite-sized} chaotic system is reminiscent of that recently observed as a consequence of ``Floquet supersymmetry''~\cite{Iadecola:2018}. 
Similarly, there, ${\mathcal{P}_S}$ can be interpreted as a projector onto 
a degenerate subspace comprising a measure zero set of nonthermal eigenstates, pinned to $ 0 $ and $ \pi $ quasi-energy modes,
which are protected by the ancillary time-reflection symmetry~\cite{Iadecola:2018}.  
By contrast, the scarring effect in our generic model is of pure dynamical origin 
that emerges in the absence of any protecting primary  symmetry, and hence does not require such tuning.

\section{emergence of quasi-conservation laws}\label{sec:WEB}
We now turn to the explanation of the SDTC dynamics via the \textit{emergence} of dynamical constraints in the form of long-lived local quasi-conservation laws. To this end, we look at the stroboscopic evolution of participation entropy,
\begin{align} \label{eq:}
{S_d (nT)=-\sum_{|i\rangle\in\mathcal{H}_d}|\langle i|\psi(nT)\rangle|^2 \log |\langle i|\psi(nT)\rangle|^2, } 
\end{align} 
starting from the CDW state $|\psi_0\rangle $, which measures the spreading of an initial wavefunction over a certain basis in the course of time.
Here the computational basis is grouped into the subspaces $ {\mathcal{H}_d} $, each of them has a fixed Hamming distance from $|\psi_0\rangle $; the distance which is defined as a minimum number of particle exchanges required to transform a specific basis into the CDW pattern. Clearly, 
at the limit of $ {\lambda/V\to\infty} $, e.g., at the exact integrable $ {V=0} $ line, 
the system possesses explicit local conservation laws over multiples of two driving periods, i.,e.,  ${[\hat{\mathcal{I}}_{i},\mathcal{U}_F^{2n}]=0}$, and applying $ \mathcal{U}_F $ to $ |\psi_0\rangle $ displaces the state into its particle-hole counterpart with the maximum  
distance, ${ d_{max}=N/2 }$. The subsequent action of $ \mathcal{U}_F $ will bring it back to itself at $ {d_{min}=0} $, closing the cycle at time $ 2T $. 
This procedure is carried out perfectly without delocalization in any intervening subspaces such that $ {S_d (nT)=0}$ for all $ d $, even at infinite time.

For a typical finite value of $ V $, however, there is no such an exact conservation. 
Nonetheless, the behavior of $ {S_d (nT)}$ shown in  Fig.~\ref{fig:D_cons}(a), suggests the emergence of long-lived quasi-conservation laws within the SDTC regime: 
the wavefunction remains almost localized and only partially leaks into the nearby sectors in the vicinity of $ d_{min} $ and $ d_{max} $.
Indeed, the SDTC dynamics does not mix different eigenstates in different mutually conserved sectors and approximately preserves the underlying local conservation laws,
i.e.,  ${[{{\tilde{\mathcal{I}}}}_{i},\mathcal{U}^{2n}_F]\sim0}$.
From this 
it infers that if we label $ {|\psi_0\rangle} $ by the set of
$ {\{\mathcal{I}_i \}} $, and prob the dynamics stroboscopically in multiples of two driving periods, it cannot strikingly evolve out to a different subspace even in the absence of explicit local conservation laws.
 
This 
bears some resemblance to the fracturing effect~\cite{Pai:2019,Pai:2019_fractonlocalization,Khemani:2019_shatter,Hudomal:2020,Yang:2019}, in which a number of
\textit{product} 
states (the so-called inert configurations~\cite{Khemani:2019_shatter,Sala:2020}) remain invariant by the dynamics and construct (exactly) localized Krylov subspaces (of dimension one), characterized by a set of \textit{state-dependent}, quasi-local integrals of motion.
It should be noted that this phenomenon does not necessary require an explicit form of fracton-like constraints, i.e., charge and dipole conservation. 
Such a constrained dynamics can also asymptotically emerge from the confinement of quasiparticle excitations
in the strongly interacting limit of some unconstrained
Hamiltonians~\cite{Yang:2019,Pai:2020_con}, specifically those with a similar form to $ H_\mathcal{D} $ in Eq.~\eqref{eq:FM1}~\cite{Tomasi:2019_frac}.
However, in contrast to the standard fracturing phenomenon where the anomalous nonthermal eigenstates have a perfect product form, here dynamical scars 
possess cat-like structure accounts for the robust period-doubling effect.

At the ergodic side of the thermalization crossover, e.g., $ {V=0.3} $, the wavefunction begins to marginally spread over increasingly distant other sectors, while still preserves its coherent oscillatory behavior during early-to-intermediate times.
However, the substantial growth of $ S_d (nT) $ eventually happens at the late times, implying the delayed onset of Floquet thermalization. 
By further increasing $ V $ deep in the thermal regime, the rapid expansion of the initial wavefunction indicates the absence of any emergent constraint, 
which in turn allows the model to thermalize faster.

The same conclusion also holds when one rearranges the computational basis $ |i\rangle $, according to the sectors characterized by total doublon density, $ {\mathcal{D}} $. Again in the SDTC regime, the wavefunction  partially delocalizes about its initial doublon sector, i.e., $ {{\mathcal{D}}=0} $, and does not explore its entire phase space (see $ {S_{\mathcal{D}}(nT)} $ at $ {V=0.1} $ shown in Fig.~\ref{fig:D_cons}(b)).
Hence, the \textit{dynamics} starting from $ |\psi_0\rangle $ is effectively restricted to approximately preserve the initial doublon number. 
The emergence of doublon conservation occurs in spite of the fact that the dominant eigenstates of the Floquet operator do not generally exhibit such a conservation law and look like the \textit{featureless} infinite-temperature states (as already mentioned in Fig.~\ref{fig:doublon_Daelta0_OBC}). 
All these results indicate that the emergence of quasi-conservation laws in the model~\eqref{eq:FM1} is of pure dynamical origin, through which an initial-state dependent, nonergodic dynamics would happen in the strongly interacting limit of a generic Floquet system.
\section{Conclusions and Discussion}\label{sec:Discussion}
In summary, we presented compelling evidence that
the quantum many-body dynamics of
strongly interacting, 
chaotic Floquet systems can exhibit anomalous discrete 
time-translation symmetry breaking, 
protected by weak ergodicity breaking. 
This breakdown is attributed to the presence of special scar subregion of the Hilbert space with anomalous long-range order,
which leads to a robust, long-lasting SDTC dynamics from a family of experimentally accessible initial states.  
The stability of dynamical scars to generic perturbations of the drive reflects
the rigidity of the SDTC response. 
We utilize the confused recurrent neural network to keep track the crossover between SDTC and fully ergodic regimes, purely from dynamics.
Such semisupervised machine learning based approach can also be amenable for classifying other partially/fully nonergodic phases~\cite{Khemani:2019_shatter,Sala:2020,Morningstar:2020}, impressively, when the exact structure of dynamical phase diagram is not known beforehand. 

The chaotic model considered in this work can be seen as a  generic deformation of the parent nonergodic model ${\mathcal{U}_{F}^{int}(\varepsilon,\lambda,\Delta)}$~\cite{Huang:2018}, which contains emergent {integrable manifold} in a wide range of its parameter space.  
By deformation of ${\mathcal{U}_{F}}$ towards ${\mathcal{U}_{F}^{int}}$, the subharmonic response will be enhanced up to a
finite time scale $ \tau_c $,
corresponding to a characteristic length
scale $\ell_{c}$,  
and not necessary being infinitely long-lived.
These features are analogous to those of generic (time-independent) scarred  Hamiltonians that are in proximity to a putative integrable \textit{point}~\cite{Khemani:2019}. 
A similar thermalization length scale can also appear in some variants of disorder-free MBL, e.g., Stark (or Bloch) MBL~\cite{Schulz:2019,vanNieuwenburg:2019} in the presence of large but finite tilts~\cite{Khemani:2019_shatter},
or quasi-MBL arising in a finite homogeneous system with drastically different energy
scales~\cite{Schiulaz:2015,Papic:2015,Yao:2016,Yarloo:2018}.	
In all mentioned cases, the life time of nonergodic dynamics looks seemingly infinite in relatively small system sizes.
Although thermalization can eventually be restored at long (but finite) time scales
in the thermodynamic limit.
This observation implies the \textit{partial} persistence of the SDTC,
in a sharp contrast to the conventional MBL-DTCs that remain absolutely stable in the limits $ {t\to\infty} $ and $ {N\to\infty} $.
It should be noted that while the observed period-doubled dynamics
in $ \mathcal{U}_F $ does not fulfill the strict definition of time crystals, it persists for much longer times than are accessible in nowadays experimental settings.
Besides, the arguments along the lines of Ref.~\onlinecite{Khemani:2019} suggest 
a tendency towards thermalization that would arise in the deformed model ${\mathcal{U}_{F}^{int}}$ (or, more generally, in any nonergodic model with a \textit{finite} distance from an exact integrable manifold/point); 
a natural tendency that may appear at \textit{larger} time/length scales than are accessible by the simulation methods.
This is the reason why in the strongly interacting regime of ${\mathcal{U}_{F}^{int}}$, $ \tau_c $ seems infinite in finite-sized systems~\cite{Huang:2018}.

However, our investigation
does not strictly rule out
the possibility of 
the existence of a genuine SDTC phase which satisfies the strict definition of time crystals in the presence of quantum chaos.
One of the most promising directions for future works is finding the signs of such an exotic DTC phase in systems exhibiting \textit{strong} fragmentation~\cite{Khemani:2019_shatter,Sala:2020,Pai:2019,Pai:2019_fractonlocalization,Moudgalya:2019,Morningstar:2020}; a phenomenon that provides a concrete (and more provable) paradigm of partially nonergodic phases instead of transient regimes. 
One can examine whether the \textit{exactly} localized subspaces of the underlying models might be amenable to harbor $ \pi $ spin-glass order persisting for an \textit{infinite} time.  
Such a study can spell out the minimal ingredients needed for realizing a true clean DTC.

Another outstanding challenge is whether the SDTC phenomenon can be reconciled using the framework of mixed phase space which has been recently extended to the realm of \textit{many-body} chaos and accounts for weak ergodicity breaking on general grounds~\cite{Michailidis:2020}. 
This standpoint is based on projecting many-body quantum dynamics into effective classical equations of motion through time-dependent variational principle (TDVP) in a restricted MPS manifold; a semiclassical approach that in general stands beyond mean-field description. 
It may be highly valuable to examine the SDTC dynamics and its stability via the aforementioned TDVP ans\"{a}tze and characterize possible deformations that increase this stability through the concept of ``quantum leakage''~\cite{Michailidis:2020}.
Such a study on the one hand gives an intuition about the notion of dynamical scars in the model~\eqref{eq:FM1},
and on the other hand sheds light on the relevant parameter regime, local observables, and initial conditions for which the SDTC behavior may be potentially observed.
\section*{Acknowledgements} 
We gratefully acknowledge A.~Vaezi, M.~Kargarian, S.~S.~Jahromi, Y.~Javanmard, and T.~L.~M. Lezama for helpful discussions surrounding this work. The authors would like to thank Sharif University of Technology for financial supports under Grant No.~G960208, and CPU time from LSPR and Cosmo HPC machines. Our numerical schemes are based on the PETSc~\cite{petsc-user-ref,petsc-efficient} and SLEPc~\cite{Hernandez:2005} libraries.
%
\end{document}